# Certain new M-matrices and their properties and applications*

R.N.Mohan[1], Sanpei Kageyama[2], Moon Ho Lee[3], and Gao Yang[4].

Sir CRR Institute of Mathematics, Eluru-534007, AP, India[1];

Graduate School of Education, Hiroshima University, Highashi-Hiroshima, 739-8524, Japan[2];

Institute of Information & Communication, Chonbuk National University, Korea[3];

Center for Combinatorics, Nankai University, Tianjin-300071, PR China[4].

Email: mohan420914@yahoo.com[1], ksanpei@hiroshima-u.ac.jp[2], moonho@chonbuk.ac.kr[3].

Abstract: The $M_n$-matrix was defined by Mohan [20] in which he has shown a method of constructing (1,-1)-matrices and studied some of their properties. The (1,-1)-matrices were constructed and studied by Cohn [5],Wang [33], Ehrlich [8] and Ehrlich and Zeller[9]. But in this paper, while giving some resemblances of this matrix with Hadamard matrix, and by naming it as M-matrix, we show how to construct partially balanced incomplete block (PBIB) designs and some regular bipartite graphs by it. We have considered two types of these M- matrices. Also we will make a mention of certain applications of these M-matrices in signal and communication processing, and network systems and end with some open problems.

## 1. Introduction

The (1,-1)-matrices have been investigated by many authors who also studied their properties. For an n×n (1,-1)-matrix having the largest possible determinant called (Hadamard's maximum determinant problem is well known) as Hadamard matrix. ], Ehrlich [8] and Ehrlich and Zeller[9] studied these binary matrices and Seifer [27] and Wang [33] studied their properties and Kahn, Kolmos and Szemeredi [17] studied the probability of these matrices to be singular. Again for the Hadamard matrix, refer to Geramita and Seberry [11], Seberry and Yamada [26]. These matrices have wide applications in the construction of codes, designs and graphs [11, 26], and of sequences that were used in signal processing (Fan and Darnell [10]). As the matrices that are being discussed here are non-orthogonal, this property of non-orthogonality is well used in image reconstruction in image analysis for details refer to Teague [31]. The concepts of orthogonal matrices, quasi orthogonal matrices by Zafarkhani [14], and non-orthogonal matrices are being used in the construction of space-time block codes by Yu hang, Yingbo Hua and Sudler, [34]. A 4×4 quasi-orthogonal code matrix is that the columns are grouped such that, a pair of columns of the different groups are orthogonal but the columns in the same group are not orthogonal, as per the definition of Jafarkhani [14]. The $M_n$-matrices were defined by Mohan [20] as the matrix constructed from the formula $M_n = (d_i \otimes d_h d_j)$ mod n by suitably defining $d_i$, $d_h$, $d_j$ and $\otimes$. By using this $M_n$-matrix pattern Vasic and Milenkovic [32, pp.1165] gave a method of construction of

---

Low-density parity check (LDPC) codes. Kageyama and Mohan [15,16] used these $M_n$-matrices for the construction of µ-resolvable and affine µ-resolvable balanced incomplete block (BIB) and PBIB designs. And in [20] two types of (1,-1)-matrices were constructed and some of their properties were studied. Now we consider those two types of matrices, i.e., when (i) n is a prime, and (ii) n+1 is a prime, and study them further.

The Hadamard matrix (H-matrix) of order n is a square matrix, which satisfies $H H' = n I_n$, where $I_n$ is the identity matrix of order n. Now while making the comparative study of the present matrix and the H-matrix we will show how best we can make use of them, to construct some symmetric partially balanced incomplete block (SPBIB) designs and some regular bipartite graphs.

In the literature that many people worked on PBIB designs, for example refer to Bose [4], Goethels and Siedel [12], Liu [18], Mohan [19], Raghavarao [23], and especially Ramanujacharyulu [25], Shrikhande [28], and Sprott [30], studied the designs with three and more associate classes. Whereas some researchers worked on symmetric partially balanced incomplete block (SPBIB) designs, for further details refer to Bose [4], Liu [18], Raghavarao [23]. For many other properties of BIB, SBIB and PBIB designs, Raghavarao [24] is a good reference.

**Definition 1.1**. A BIB design is an arrangement of v symbols in b blocks of size k each such that every symbol occurs in r blocks and each pair of symbols occurs in λ blocks. A BIB design satisfies the following conditions:

$$v r = b k, \lambda (v-1) = r (k-1), b \geq v.$$

A BIB design with v = b is said to be symmetric.

Bose and Nair [3] defined a PBIB design in which some pairs of symbols do not occur in a constant number of times as in a BIB designs. For defining a PBIB design we require the concept of an association scheme with m associate classes.

**Definition 1.2**. Given v symbols 1, 2, …, v, a relation, which satisfies the following conditions is said to form an association scheme with m associate classes.

1. Any two symbols are either $1^{st}$, $2^{nd}$,…, or $m^{th}$ associates. The relation of association is symmetric, that is, if the symbol α is the $i^{th}$ associate of the symbol β, then β is the $i^{th}$ associate of α.
2. Each symbol α has $n_i$ $i^{th}$ associates, the number $n_i$ being independent of α.
3. If any two symbols α and β are $i^{th}$ associates then the number of symbols that are $j^{th}$ associates of α and $\ell^{th}$ associates of β is $p^i_{j\ell}$ and is independent of the pair of $i^{th}$ associates α and β.

The numbers v, $n_i$ (i = 1, 2, …, m), $p^i_{j\ell}$, (i,j,$\ell$ =1, 2,.., m) are called parameters of the association scheme.

Given an association scheme for the v symbols we define a PBIB design as follows:

**Definition 1.3.** When we have an association scheme with m associate classes, an m-associate PBIB design with parameters v, b, r, k, $\lambda_i$ (i =1, 2, …, m) can be defined if the v symbols are arranged in b blocks of size k (< v) such that
1. every symbol occurs at most once in a block,
2. every symbol occurs exactly in r blocks,
3. if two symbols α and β are $i^{th}$ associates, then they occur together in $\lambda_i$ blocks, the number $\lambda_i$ being independent of the particular pair of the $i^{th}$ associates α and β.

The numbers v, b, r, k, $\lambda_i$ (i= 1, 2, …, m) are called parameters of the design.

A PBIB design satisfies the following conditions:

$$vr = bk, \ \sum_{i=1}^{m} n_i = v-1, \ \sum_{i=1}^{m} n_i \lambda_i = r(k-1), \ \sum_{\ell=1}^{m} p^i_{j\ell} = n_j - \delta_{ij},$$

where $\delta_{ij}$ is the Kronecker delta taking the value 1 when i = j, otherwise 0.

**Definition 1.4.** Let G = (V, E), where V is a set of vertices and E is a set of edges, joining any two vertices, be called as **graph.**

The number of edges passing through a vertex is called its **valence.** In a graph if the valence of each of its vertices has the same constant then it is said to be **regular**.

**Definition 1.5.** If the vertex set V has two complementary sets $V_1$, $V_2$ such that each edge of the graph has one end in $V_1$ and the other end in $V_2$, then the graph is called as **bipartite graph.**

Mohan in [20], defined an $M_n$-matrix $=(a_{ij})$, with $a_{ij}$ = ($d_i \otimes d_h d_j$) mod n by suitably defining $d_i$, $d_h$, $d_j$ and $\otimes$, which is given below.

**Definition 1.6.** *When n is a prime an $M_n$-matrix $(a_{ij})$ is defined as a matrix obtained from*

$a_{ij} = 1 + (i-1)(j-1) \mod n, i, j = 1, 2, ..., n$. *This is an $n \times n$ symmetric matrix.*

This has been used in the construction of graphs (refer to [20]) called as $M_n$-graphs, which are defined as follows:

**Definition 1.7.** *Given an $M_n$-matrix and $C_k$ be its columns, which are numbered as 1,2,…,n and $a_{ij}$'s be the elements of it, and then let $V_1 = \{C_k\}$, $V_2 = \{a_{ij}\}$ be the two sets of the vertices. Now there is an edge $\alpha_{ijk}$ iff $a_{ij}$ is in $C_k$. Then the graph is ($V_1$, $V_2$, $\alpha_{ijk}$), where i,j,k = 1,2,…,n, is called as $M_n$-graph.*

As an extension of these concepts, we define an M-matrix as follows:

**Definition 1.8.** When n is a prime, consider the matrix of order n obtained by the equation $M_n = (a_{ij})$, where $a_{ij} = 1 + (i-1)(j-1)$ mod n, i, j = 1, 2, …, n. In the resulting matrix retain 1 as it is and substitute -1's for odd numbers and +1's for even numbers. (We can substitute 1 for odd numbers and -1 for even numbers, in that case change of sign occurs in its determinant). Let the resulting matrix M be called as M-matrix of Type I. This is a n×n symmetric matrix.

The M-matrix of Type II is obtained by the equation $a_{ij} = (i.j)$ mod n+1, where n+1 is a prime and i,j = 1,2,…n. In this matrix since each row or column has n elements whereas n is even, 1 to n elements do come in all the columns and rows, and each element comes once in each row and each column. In the resulting matrix substitute 1 for even numbers and -1 for odd numbers and also for 1, ( or 1 for odd numbers keeping the 1 in the matrix as 1 itself and -1 for even numbers). Then this resulting matrix M is called M-matrix of Type II. Each row (column) consists of an equal number of +1's and -1's numbering to n/2. This is also a n×n symmetric matrix.

We discuss these two types of matrices, while giving examples for their constructions and applications of these matrices in the later sections.

Take an M-matrix of Type I and the Hadamard matrix (H-matrix).Then we see some resemblances and differences between them.

1. In both the M-matrix and the H-matrix the first rows and the first columns consist of unities only. Both are symmetric matrices.
2. In both the matrices all the elements are either 1 or -1.
3. In the H-matrix the row sums (except the first row) are all zeros and in the M-matrix each of the row sum (except the first row) is one. When in the H-matrix the row sums are some constant then it is called a regular Hadamard matrix, but in the case of the M-matrix it is 1.
4. The H-matrix may exist for n = 2 or n ≡ 0 mod 4, but the M-matrix exists for any prime n.
5. The H-matrices are used in the constructions of codes, graphs and designs, refer to Seberry and Yamada [29]. In this paper it will be shown that the M-matrix also can be used in the construction of certain designs and graphs.
6. If $H_n$ and $H_m$ are two Hadamard matrices then their Kronecker product $H_n \otimes H_m$ is also a Hadamard matrix, but it is **not** so in the case of the M-matrices.
7. From an H-matrix one can construct a symmetric BIB design and it will be shown that from an M-matrix one can construct a symmetric PBIB design.
8. The H-matrices are orthogonal matrices and these M-matrices are non-orthogonal.

There are wide applications of matrices/ BIB designs and their associated graphs in information and communication systems and in network systems. For further details refer to Colbourn [6], Colbourn, Dinitz, Stinson [7], Aupperle and Meyer [1], Skillicorn [29], Mohan and Kulkarni [21], Bhuyan, and Agarwal [2], Hawkes [13], and Ngoc Chi Nguyen, Nhat Minh Dinh Vo and Sung young Lee [22].

## 2. Statements

As these two types of M-matrices are non-orthogonal, we define the orthogonal numbers for them as follows:

**Definition 2.1.** The **orthogonal number** of a given M-matrix with entries ± 1 is defined as sum of the products of the corresponding numbers in two given rows of the matrix (called an inner product of the rows). Consider any two rows $R_l = (r_1, r_2, ..., r_n)$ and $R_m = (s_1 s_2, ..., s_n)$ and then the orthogonal number denoted by 'g' can be defined as $g = (R_l R_m) = \sum_{i=1}^{n} r_i s_i$.

**The M-matrix of Type I.** The elements of the principal diagonal of this matrix has a special property, i.e. if $D_n$ is the principal diagonal of the M-matrix of order n, then we get its elements by $(x^2-2x+2) \mod n$. Thus for example $D_3 = (1\ \mathbf{2}\ 2)$, $D_5 = (1\ \mathbf{2}\ 5\ 5\ 2)$ and $D_7 = (1\ \mathbf{2}\ 5\ 3\ 3\ 5\ 2)$, leaving the first element the remaining (n-1)/2 elements just repeat in the reverse order. It is a modular property.

**Proposition 2.1.** In a given M-matrix, in each of its rows and columns, the number of +1's are (n+1)/2 and the number of -1's are (n-1)/2.

**Proof.** Since n is a prime and as the first row and the first column have unities and hence in each row (column) the first element is 1 and among the remaining (n-1) places of each row(column) there occur each of 2, 3, …, n exactly once. And among these n-1 elements half of them are even numbers and the other half of them are odd numbers. Consequently as we replace even numbers by +1's and odd numbers by -1's, we get that (n+1)/2 elements are +1's (as we add the first element 1 also), and (n-1)/2 elements are -1's.

**Proposition 2.2.** In an M-matrix of order n (prime), the orthogonal number between any two distinct rows $R_i$ and $R_j$ (where $i \neq j$) is given by 4k+2-n, where k is the number of unities in the selected set and $0 \leq k \leq \frac{n-1}{2}$.

**Proof.** Let $R_i$, and $R_j$ be the two given rows in the M-matrix. We have to calculate their inner product $\langle R_i, R_j \rangle$, where $i \neq j$ and $i \neq 1$. By elementary transformations we can make the row $R_i$ with the first $\frac{n+1}{2}$ elements as 1's and the next $\frac{n-1}{2}$ elements as -1's. The orthogonal numbers of the matrix remains invariant by such elementary transformations. Now consider the other row $R_j$, which has n elements. First element is 1 and the remaining n-1 elements divide them into two halves with $\frac{n-1}{2}$ elements each. They can be depicted as follows:

$$R_i = \begin{pmatrix} \overset{\frac{n-1}{2}\,elements}{(1)(1\,1\,1\,1\ldots1\,1\,1\,1\ldots1)} & \overset{\frac{n-1}{2}\,elements}{(-1-1-1-1\ldots-1-1-1-1)} \end{pmatrix}$$

$$R_j = \begin{pmatrix} (1)(1\text{-}1\text{-}1\ldots\text{-}1\text{-}1\ldots1\,1) & (1\,1\ldots1\text{-}1\,1\text{-}1\text{-}1\text{-}1\text{-}1\,1\ldots1\,1) \\ \text{let +1's be k in number hence -1's are } \frac{n-1}{2}\text{-}k & \text{+1's are } \frac{n-1}{2}\text{-}k \text{ and -1's are k in number} \end{pmatrix}$$

Now we evaluate the formula for the orthogonal number. Let k be the number of +1's, excluding the first element +1. In the two given rows (i) The first elements of the two rows 1 coincides with 1, in the first two sets of the given rows (ii) the k number of +1's of the row $R_j$ correspond with k number of +1's in $R_i$. Now in the second sets of the two given rows (iii) $\frac{n-1}{2}$-k number of +1's of $R_j$ correspond with the same number of -1's of $R_i$ and (iv) the k number of –1's of $R_j$ correspond with the same of -1's in $R_i$. Hence we get

$$g = \langle R_i, R_j \rangle = (1\times1) + (k\times1\times1) + ((\frac{n-1}{2} - k)\times1\times(-1)) + ((\frac{n-1}{2} - k)\times(1)\times(-1)) + k(-1)(-1) = 4k + 2 - n.$$

If we see for different values of k, when k = 0 then g = 2-n, k=1, g = 6-n, and so on. And when k = (n-1)/2 then g = n. Thus there should be $\frac{n-1}{2} + 1 = \frac{n+1}{2}$ orthogonal numbers in an M-matrix. □

**Result. 2.1.** We have $\langle R_1, R_j \rangle = 1, \langle R_i, R_i \rangle = n$ and $\langle R_{2+i} R_{n-i} \rangle = 2 - n, i = 0, 1, 2, \ldots, \frac{n-3}{2}$.

**Proof.** We have $\langle R_1, R_j \rangle = 1$, since $R_1$ consists of all 1's and the inner product with any row consequently is 1. This situation has not been counted in the formula 4k+2-n. And for $\langle R_i, R_i \rangle = n$, the inner product of any row with itself, gives out n only. Besides when k = $\frac{n-1}{2}$, then also we get $\langle R_i, R_j \rangle = n$. They are called trivial orthogonal numbers. The third equality is trivial by proposition 2.2. □

**Note.2.1** There are some orthogonal numbers which occur in pairs only. They are called orthogonal pairs, which we discuss in later sections.

**Note 2.2.** There are certain exceptions depending on the coincidences. Some times the theoretically obtained $g_i$'s may not exist in numerical problems. For example when n = 11, orthogonal numbers should be -9,-5,-1,3,7,11, but in the numerical problem we get these orthogonal numbers as -9,-1,3.

The orthogonal numbers -5 and 7 do not exist. Of course 1 and 11 will be there, which are trivial orthogonal numbers, when $\langle R_1, R_j \rangle = 1, \langle R_i, R_i \rangle = 11$ are considered. Besides the trivial orthogonal number this 1 may again exist as an orthogonal number in the list, like in the case of n = 5, and k = 1, g = 1. □

**Proposition 2.3.** In an M-matrix the sum of the orthogonal numbers is $\frac{n+1}{2}$. If 1 of $\langle R_1, R_j \rangle = 1$ is also added then it will be $\frac{n+1}{2} + 1 = \frac{n+3}{2}$.

**Proof.** When $g_i$'s are orthogonal numbers of the given M-matrix, $\sum_{i=1}^{\frac{n+1}{2}} g_i = \sum_{k=0}^{n-1/2} 4k + 2 - n = (n+1)/2$, which is an arithmetic progression with AM as 4. The orthogonal number obtained from the first row $R_1$ with any other row is not considered in this. Hence the proof. □

Note that in considering the sum, all possible orthogonal numbers by the formula are taken.

**Note 2.3.** If $M_1$ and $M_2$ are two given M-matrices of order m and n respectively, then the Kronecker product $M_1 \otimes M_2$ is not an M-matrix. Since this M-matrix exists for n being prime, when two matrices of order m, n are taken then their product m×n ceases to be a prime number and hence the resulting matrix is not an M-matrix.

It is not possible to derive any other M-matrix from either the given M-matrix or from the given set of M-matrices by any other means because the order of the resulting matrix ceases to be a prime number.

**Result 2.2.** For the given M-matrix | M |     = -4 if n = 3
                                                                 = 0 if n ≥ 5.

Proof. For n = 3 it is trivial. But, when n ≥ 5, the matrix consists of pairs of rows with just opposite signs, and by leaving the first row, consider $R_2 + R_n$. It will be a vector (2 0 0 …0) and $R_3 + R_{n-1}$ will also be the same vector (2 0 0 .. 0). In any matrix, if any two rows are equal then its determinant is zero, i.e. $|M| = 0$.

Now we consider $MM'$, which is an $n \times n$ symmetric matrix. By the properties of orthogonal numbers discussed above we have $\langle R_1, R_j \rangle = 1, \langle R_i, R_i \rangle = n$ and $\langle R_{2+i} R_{n-i} \rangle = 2 - n$. The elements in the principal diagonal are n, rest of the n-1 elements in the first row and the fist column are unities. By considering all the other elements in this matrix, they are some orthogonal numbers, which can not be fixed theoretically. The number of orthogonal numbers is (n+1)/2, and in a row of n elements (n-1) orthogonal numbers have to be listed and hence some orthogonal numbers repeat, when all the possible orthogonal numbers exist. Otherwise, if some do not exist then the available numbers will repeat in the row.

**Proposition 2.4.** The existence of an M-matrix of order n, where n is a prime, implies the existence of a SPBIB design with parameters v = b = n-1, r = k = (n-1)/2, $\lambda_i$ vary from 0 to (n-3)/2.

$n_i$, $p^i_{jk}$ can not be said theoretically, but can be evaluated in the numerical problems.

**Proof.** Given an M-matrix of order n, where n is a prime. Leaving the first row and the first column we get a matrix of order (n-1). Let the matrix is of the form

$$M = \begin{bmatrix} 1 & \underline{\Uparrow}_{1,n-1} \\ \underline{\Uparrow}_{n-1,1} & D \end{bmatrix},$$

where $\underline{\Uparrow}$ is the matrix of given order with all 1's. Let $D_1$ be the matrix obtained from D, by replacing -1's by 0's. Then $D_1$ is the incidence matrix of a PBIB design with parameters v = n-1 = b, r = k = (n-1)/2. Regarding $\lambda_i$ values in the given two rows $R_i$ and $R_j$ if they have exactly opposite signs then $\lambda_i$ value will be zero. In the given two rows $R_i$ and $R_j$ if they have x coincidences then $\lambda_i$ = x. Then in each row the number of 1's are (n-1)/2 and the number of zeros is also (n-1)/2. All (n-1)/2 number of 1's cannot coincide with (n-1)/2 number of 1's of another row, if so the two rows are identical, which is not possible. Hence at the maximum there may be (n-1)/2 – 1 = (n-3)/2 coincidences. Hence $\lambda_i$ values vary from 0 to (n-3)/2. Consequently $n_i$'s and $p^i_{jk}$'s can be evaluated□

**Proposition 2.5.** The existence of an M-matrix of Type I implies the existence of a regular bipartite graph with parameters V = 2n ($V_1$= n, $V_2$=n), E = 2n and valence is $\frac{n-1}{2}$.

Procedure for the construction: Given an M-matrix, treating -1's as 0's and by considering the resulting matrix as an adjacency matrix of a graph with elements as one set and the columns as another set, if an element $a_{ij}$ is in column '$c_k$' then ($a_{ij}$,$c_k$) will be an edge, otherwise not. We have in the graph obtained V = 2n, ($v_1$ = n, $v_2$ = n) and its valence is $\frac{n-1}{2}$, since each column and each row has $\frac{n-1}{2}$ unities. Hence we will get a regular bipartite graph.

**Example 2.1** Take n = 5. From $M_n$ = ($a_{ij}$), where $a_{ij}$ = 1 + (i-1)(j-1) mod n, i, j = 1, 2, 3,4,5.

$$M_n = \begin{bmatrix} 11111 \\ 12345 \\ 13524 \\ 14253 \\ 15432 \end{bmatrix}.$$

Now by substituting for even number 1 and for odd number -1 and keeping 1 as it is we get

$$M = \begin{bmatrix} 1 & 1 & 1 & 1 & 1 \\ 1 & 1 & -1 & 1 & -1 \\ 1 & -1 & -1 & 1 & 1 \\ 1 & 1 & 1 & -1 & -1 \\ 1 & -1 & 1 & -1 & 1 \end{bmatrix}.$$

Leaving the first row and the first column of unities we have

$$M = \begin{bmatrix} 1 & -1 & 1 & -1 \\ -1 & -1 & 1 & 1 \\ 1 & 1 & -1 & -1 \\ -1 & 1 & -1 & 1 \end{bmatrix}.$$

Now consider all -1 s as zeros then we get the matrix N as

$$N = \begin{bmatrix} 1010 \\ 0011 \\ 1100 \\ 0101 \end{bmatrix}.$$

Consider this matrix N as an incidence matrix of a design, which yields a PBIB design with parameters v = 4 = b, r = k = 2, $\lambda_1$ = 1, $\lambda_2$ = 0, which is a two-associate $L_2$ PBIB design where the other values are given by

$p_{11}^1 = 0, p_{12}^1 = 0, p_{21}^1 = 0, p_{22}^1 = 2, p_{11}^2 = 0, p_{12}^2 = 1, p_{21}^2 = 1, p_{22}^2 = 0, n_1 = 1, n_2 = 2$.

If we consider this matrix N as the adjacency matrix of a graph then we get

```
     5   6   7   8
1    1   0   1   0
2    0   0   1   1
3    1   1   0   0
4    0   1   0   1
```

a regular bipartite graph as follows:

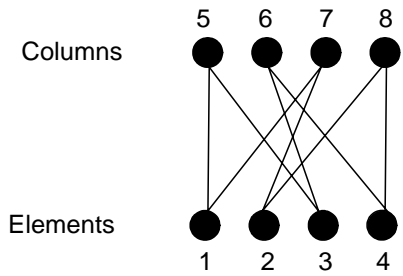

with parameters (4,4;8). These types of graphs form a new family of graphs, which are highly usable in routing problems of salesmen, transportation or communication problems. Suppose there are n nodes (points) say ($a_1, a_2, .., a_n$). Consider any point $a_1$ as a source node and then all the remaining (n-1) points are destination nodes. By starting from the source node the message has to be passed on to any other node with the condition that it should not be passed on to the nodes of same set twice continuously at any time, that is alternatively the message should be passed on to the nodes of different sets $V_1$ and $V_2$ with an exception to the source node at the final stage if needed. Iteratively this goes on in succession and reaches back to the first source node. All the points should be covered. And for each node have two ways of going in and coming out only. Hence each node has only two edge connections. These types of graphs are specifically useful, in passing information very confidentially and routing the messenger to go without informing the persons in the vicinity, but covering all the members (nodes) and reporting back to the source node. The maximum number of lines required is n. But the minimum number of lines required varies as per the task. And how many ways are there for having this function to be performed with the minimum number of hops can be evaluated. This is a richly connected graph to act as a network. For further details refer to Mohan and Kulkarni [22], Bhuyan, Agarwal [2], Hawkes [13], and Ngoc Chi Nguyen, Nhat Minh Dinh Vo and Sung Young Lee [23].

**Example 2.2.** Similarly when n = 7 we can construct a 3-associate class PBIB design. By taking $a_{ij} = 1+(i-1)(j-1)$ mod n, where i,j = 1,2,3,4,5,6,7, we get the $M_n$-matrix as

$$\begin{bmatrix} 1 & 1 & 1 & 1 & 1 & 1 & 1 \\ 1 & 2 & 3 & 4 & 5 & 6 & 7 \\ 1 & 3 & 5 & 7 & 2 & 4 & 6 \\ 1 & 4 & 7 & 3 & 6 & 2 & 5 \\ 1 & 5 & 2 & 6 & 3 & 7 & 4 \\ 1 & 6 & 4 & 2 & 7 & 5 & 3 \\ 1 & 7 & 6 & 5 & 4 & 3 & 2 \end{bmatrix}.$$

Now by substituting for even numbers +1 and for odd numbers -1 and keeping 1 as it is we get the M-matrix of Type I as

$$\begin{bmatrix} 1 & 1 & 1 & 1 & 1 & 1 & 1 \\ 1 & 1 & -1 & 1 & -1 & 1 & -1 \\ 1 & -1 & -1 & -1 & 1 & 1 & 1 \\ 1 & 1 & -1 & -1 & 1 & 1 & -1 \\ 1 & -1 & 1 & 1 & -1 & -1 & 1 \\ 1 & 1 & 1 & 1 & -1 & -1 & -1 \\ 1 & -1 & 1 & -1 & 1 & -1 & 1 \end{bmatrix}.$$

By leaving the first row and first column of unities and treating -1's as 0's in the resulting matrix we get an incidence matrix of a PBIB design, whose incidence matrix is given by

$$\begin{bmatrix} 101010 \\ 000111 \\ 100110 \\ 011001 \\ 111000 \\ 010101 \end{bmatrix}.$$

Hence this gives the solution of the concerned PBIB design, this has been dealt with in the example 2.5 later.

**Example 2.3.** Similarly, when n = 11 we get the $M_n$-matrix as

| 0 | 1 | 2 | 3 | 4 | 5 | 6 | 7 | 8 | 9 | 10 |
|---|---|---|---|---|---|---|---|---|---|----|
| 1 | 1 | 1 | 1 | 1 | 1 | 1 | 1 | 1 | 1 | 1 |
| 1 | 2 | 3 | 4 | 5 | 6 | 7 | 8 | 9 | 10 | 11 |
| 1 | 3 | 5 | 7 | 9 | 11 | 2 | 4 | 6 | 8 | 10 |
| 1 | 4 | 7 | 10 | 2 | 5 | 8 | 11 | 3 | 6 | 9 |
| 1 | 5 | 9 | 2 | 6 | 10 | 3 | 7 | 11 | 4 | 8 |
| 1 | 6 | 11 | 5 | 10 | 4 | 9 | 3 | 8 | 2 | 7 |
| 1 | 7 | 2 | 8 | 3 | 9 | 4 | 10 | 5 | 11 | 6 |
| 1 | 8 | 4 | 11 | 7 | 3 | 10 | 6 | 2 | 9 | 5 |
| 1 | 9 | 6 | 3 | 11 | 8 | 5 | 2 | 10 | 7 | 4 |
| 1 | 10 | 8 | 6 | 4 | 2 | 11 | 9 | 7 | 5 | 3 |
| 1 | 11 | 10 | 9 | 8 | 7 | 6 | 5 | 4 | 3 | 2 |

We can construct M-matrix from this as

|   | 1 | 2 | 3 | 4 | 5 | 6 | 7 | 8 | 9 | 10 | 11 |
|---|---|---|---|---|---|---|---|---|---|----|----|
| 1 | 1 | 1 | 1 | 1 | 1 | 1 | 1 | 1 | 1 | 1 | 1 |
| 2 | 1 | -1 | 1 | -1 | 1 | -1 | 1 | -1 | 1 | -1 | 1 |
| 3 | 1 | 1 | 1 | 1 | 1 | 1 | -1 | -1 | -1 | -1 | -1 |
| 4 | 1 | -1 | 1 | -1 | -1 | 1 | -1 | 1 | 1 | -1 | 1 |
| 5 | 1 | 1 | 1 | -1 | -1 | -1 | 1 | 1 | 1 | -1 | -1 |
| 6 | 1 | -1 | 1 | 1 | -1 | -1 | 1 | 1 | -1 | -1 | 1 |
| 7 | 1 | 1 | -1 | -1 | 1 | 1 | -1 | -1 | 1 | 1 | -1 |
| 8 | 1 | -1 | -1 | 1 | 1 | 1 | -1 | -1 | -1 | 1 | 1 |
| 9 | 1 | 1 | -1 | 1 | 1 | -1 | 1 | -1 | -1 | 1 | -1 |
| 10 | 1 | -1 | -1 | -1 | -1 | -1 | 1 | 1 | 1 | 1 | 1 |
| 11 | 1 | 1 | -1 | 1 | -1 | 1 | -1 | 1 | -1 | 1 | -1 |

$\langle R_i, R_j \rangle$ is the inner product of rows $R_i$ and $R_j$. Then we have

$\langle R_1 R_j \rangle = 1$, where $j = 1, 2, ..., 11$ and $\langle R_i R_i \rangle = 11$, which are trivial orthogonal numbers.

$\langle R_2, R_3 \rangle = \langle R_2, R_7 \rangle = \langle R_2, R_8 \rangle = \langle R_2, R_9 \rangle = \langle R_3, R_4 \rangle = \langle R_3, R_5 \rangle = \langle R_3, R_6 \rangle = \langle R_4, R_7 \rangle = \langle R_4, R_8 \rangle = \langle R_4, R_{11} \rangle$

$= \langle R_5, R_9 \rangle = \langle R_5, R_{11} \rangle = \langle R_6, R_8 \rangle = \langle R_6, R_9 \rangle = \langle R_6, R_{11} \rangle = \langle R_7, R_{10} \rangle = \langle R_8, R_{10} \rangle = \langle R_9, R_{10} \rangle = \langle R_{10}, R_{11} \rangle = -1$

$\langle R_2, R_4 \rangle = \langle R_2, R_5 \rangle = \langle R_2, R_6 \rangle = \langle R_2, R_{10} \rangle = \langle R_3, R_7 \rangle = \langle R_3, R_8 \rangle = \langle R_3, R_9 \rangle = \langle R_3, R_{11} \rangle = \langle R_4, R_5 \rangle = \langle R_4, R_6 \rangle = \langle R_4, R_{10} \rangle$

$= \langle R_4, R_{10} \rangle = \langle R_5, R_6 \rangle = \langle R_5, R_{10} \rangle = \langle R_6, R_{10} \rangle = \langle R_7, R_8 \rangle = \langle R_7, R_9 \rangle = \langle R_7, R_{11} \rangle = \langle R_8, R_9 \rangle = \langle R_8, R_{11} \rangle = \langle R_9, R_{611} \rangle = 3$

$\langle R_2, R_{11} \rangle = \langle R_3, R_{10} \rangle = \langle R_4, R_9 \rangle = \langle R_5, R_8 \rangle = \langle R_6, R_7 \rangle = -9$.

*In this the orthogonal numbers are -1,3,-9 only. But as per our formula g=4k+2-n, we get the orthogonal numbers as -9,-5,-1,+3,+7,+11. Because when the first set with +1 selected in $R_i$ and in $R_j$ there no row in it having only one +1, i.e., k = 1 does not exist, and hence k = 4 also does not exist. And consequently the pair of orthogonal numbers -5, and 17 are missing. The orthogonal number $\langle R_1 R_j \rangle = 1$ exists.*

By leaving the first row and first column of unities, and by taking -1's as 0's we get the incidence matrix of an SPBIB design as

|   | 1 | 2 | 3 | 4 | 5 | 6 | 7 | 8 | 9 | 10 |
|---|---|---|---|---|---|---|---|---|---|----|
| 1 | 1 | 0 | 1 | 0 | 1 | 0 | 1 | 0 | 1 | 0 |
| 2 | 0 | 0 | 0 | 0 | 0 | 1 | 1 | 1 | 1 | 1 |
| 3 | 1 | 0 | 1 | 1 | 0 | 1 | 0 | 0 | 1 | 0 |
| 4 | 0 | 0 | 1 | 1 | 1 | 0 | 0 | 0 | 1 | 1 |
| 5 | 1 | 0 | 0 | 1 | 1 | 0 | 0 | 1 | 1 | 0 |
| 6 | 0 | 1 | 1 | 0 | 0 | 1 | 1 | 0 | 0 | 1 |
| 7 | 1 | 1 | 0 | 0 | 0 | 1 | 1 | 1 | 0 | 0 |
| 8 | 0 | 1 | 0 | 0 | 1 | 0 | 1 | 1 | 0 | 1 |
| 9 | 1 | 1 | 1 | 1 | 1 | 0 | 0 | 0 | 0 | 0 |
| 10 | 0 | 1 | 0 | 1 | 0 | 1 | 0 | 1 | 0 | 1 |

This is a 3-associate SPBIB design with parameters v = b = 10, r = k = 5, $\lambda_1 = 0$, $\lambda_2 = 2$, $\lambda_3 = 3$, $n_1 = 1$, $n_2 = 4$, $n_3 = 4$,

$$P_1 = \begin{bmatrix} p^1_{11}=0 & p^1_{12}=0 & p^1_{13}=0 \\ p^1_{21}=0 & p^1_{22}=0 & p^1_{23}=4 \\ p^1_{31}=0 & p^1_{32}=4 & p^1_{33}=0 \end{bmatrix}, P_2 = \begin{bmatrix} p^2_{11}=0 & p^2_{12}=0 & p^2_{13}=1 \\ p^2_{21}=0 & p^2_{22}=0 & p^2_{23}=3 \\ p^2_{31}=1 & p^2_{32}=3 & p^2_{33}=0 \end{bmatrix}, P_3 = \begin{bmatrix} p^3_{11}=0 & p^3_{12}=1 & p^3_{13}=0 \\ p^3_{21}=1 & p^3_{22}=3 & p^3_{23}=0 \\ p^3_{31}=0 & p^3_{32}=0 & p^3_{33}=3 \end{bmatrix}.$$

By considering the above incidence matrix as an adjacency matrix of a graph, which is called the M-graph that can be drawn as follows:

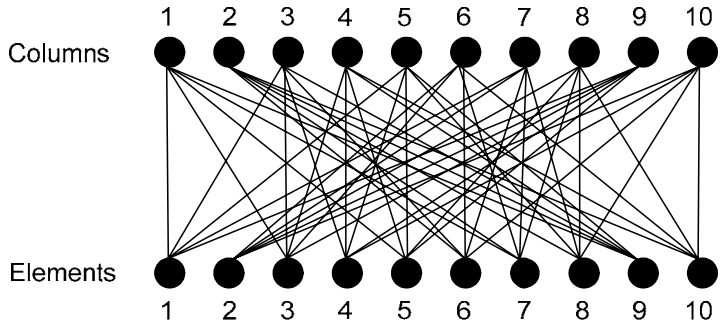

**M-matrix of Type II.** In this we consider the case where n + 1 as a prime, by considering the equation of the form $M_n = \{a_{ij}\}$, where $a_{ij} = (i \cdot j) \mod (n+1)$, where i, j = 1, 2, …, n.

The elements of the principal diagonal of this matrix has a property, i.e. if $D_n$ is the principal diagonal of the M-matrix of order n, (where n+1 is a prime), of second type then we get its elements by $x^2 \mod n+1$, quadratic residue mod n+1 elements. Thus for example $D_2 = (1\ 1)$, $D_4 = (1,4,\mathbf{4,1})$, $D_6 = (1,4,2,\mathbf{2,4,1})$. Here n/2 elements repeat in the reverse order.

**Proposition 2.6.** In an M-matrix of Type II, the orthogonal number between any two rows $R_i$ and $R_j$ where $i \neq j$ is given by 4k-n, where k is the number of unities in the selected set, where $0 \leq k \leq \dfrac{n}{2}$.

**Proof.** Let $R_i$, and $R_j$ be the two given rows in the M-matrix. We have to calculate their inner product $\langle R_i, R_j \rangle$, where $i \neq j$. By elementary transformations we can make the row $R_i$ with the first $\dfrac{n}{2}$ elements as 1's and the next $\dfrac{n}{2}$ elements as -1's. The orthogonal numbers of the matrix remains invariant by such elementary transformations. Now consider the other row $R_j$, which has n elements. Let these n elements be divided into two halves with $\frac{n}{2}$ elements each. They can be depicted as follows:

$$R_i = \underbrace{(11111...1111...1)}_{n/2 \text{ elements}} \quad \underbrace{(-1\ -1-1-1...-1-1-1-1)}_{n/2 \text{ elements}}$$

$$R_j = \underbrace{(11\text{-}1\text{-}1..\text{-}1\text{-}1..11)}_{\text{let }+1\text{'s be k in number hence }-1\text{'s are }\frac{n}{2}-k} \quad \underbrace{(11..1\text{-}11\text{-}1\text{-}1\text{-}1\text{-}11..11)}_{+1\text{'s are }\frac{n}{2}-k \text{ and }-1\text{'s are k in number}}$$

Now we evaluate the formula for the orthogonal number. In the first two sets of the two rows (i) the k number +1's of the first set of $R_j$ correspond with k number of +1's in the first set of $R_i$. (ii) $\frac{n}{2}$-k number of -1's of the first set of $R_j$ correspond with the same number of -1's of the first set of $R_i$. Now in the second sets of the two rows (iii) $\frac{n}{2}$-k number of +1's of the second set of $R_j$ correspond with the same number of -1's of the second set of $R_i$ and (iv) the k number of −1's of $R_j$ correspond with same of -1's in $R_i$. Hence we get

$$g = \langle R_i, R_j \rangle = ((k \times 1 \times 1) + ((\frac{n}{2} - k) \times (-1) \times (1)) + ((\frac{n}{2} - k) \times (1) \times (-1)) + k(-1)(-1) = 4k - n.$$

If we see for different values of k, when k = 0 then g = -n, k=1, g = 4-n, and so on when k = n/2 then g = n. Thus there should be n/2 +1 = (n+2)/2 orthogonal numbers in an M-matrix of the second type.

**Proposition 2.7.** In an M-matrix the sum of the orthogonal numbers is $\frac{n+1}{2}$.

**Proof**. If $g_i$'s are the orthogonal numbers then their sum is $\sum_{i=1}^{\frac{n+2}{2}} g_i = \sum_{k=0}^{n/2} (4k - n) = 0$. It is an AP with AM as 4, and hence the proof.

If $M$ is the M-matrix of Type II let $M'$ be its transpose then the matrix $MM'$ is also a symmetric n×n matrix, then the elements in its principal diagonal are n only as $\langle R_i, R_i \rangle = n$, which is a trivial orthogonal number. It has no second trivial orthogonal number as in the case of Type I, and the rest of its elements are the orthogonal numbers.

**Note 2. 4.** In these orthogonal numbers there exist orthogonal pairs, which either exist together or do not exist at all. The pair formation depends on the number of +1's and number of -1's in a selected set of a row. The first number of the pair occurs when the first set of a row $R_j$ is considered in which the number of +1's is $\theta_1$ and the number of -1's are $\theta_2$, and the second number of the pair occurs when the first set of a row $R_j$ is considered in which the number of -1's is $\theta_1$ and the number of +1's are $\theta_2$, where $\theta_1 + \theta_2 = \frac{n-1}{2}$ in the Type I, and $\theta_1 + \theta_2 = \frac{n}{2}$ in the Type II. For example when n = 11, $\theta_1=1$, and $\theta_2 = 4$ then -5 is the orthogonal number and again when $\theta_2=1$, and $\theta_1 = 4$ then 7 is the orthogonal number, both of these (-5,7) form an orthogonal pair, which do not occur. In it the second orthogonal pair is -9 and 11 and the third orthogonal pair is -1 and 3. In a given M-matrix of either of the types, 1 may be an orthogonal number and also $\langle R_1 R_j \rangle = 1$, where j = 1,2,…, $\frac{n-1}{2}$ in Type I. In Type II there exists **no** row with all unities satisfying and $\langle R_1 R_j \rangle = 1$.

**Result 2.3.** In an M-matrix of type I the sum of the numbers in an orthogonal pair is 2, and in type II it is 0.

**Proof.** In the M-matrix of Type I the orthogonal numbers are obtained from the formula 4k+2-n. Suppose the orthogonal pair exists for k = $\theta_1$ and k = $\theta_2$, where $\theta_1 + \theta_2 = \frac{n-1}{2}$. So we have the sum of the two orthogonal numbers in a pair = (4 $\theta_1$+2-n) + (4 $\theta_1$+2-n) = 4($\theta_1$+ $\theta_2$)+4-2n = 4 $\frac{n-1}{2}$+4-2n = 2. In the M-matrix of Type II the orthogonal numbers are obtained from the formula 4k-n. Suppose the orthogonal pair exists for k = $\theta_1$ and k = $\theta_2$, where $\theta_1 + \theta_2 = \frac{n}{2}$. So we have the sum of the two orthogonal numbers in a pair = (4 $\theta_1$-n) + (4 $\theta_1$-n) = 4($\theta_1$+ $\theta_2$)-2n = 4 $\frac{n}{2}$-2n = 0. □

**Proposition 2.8.** The existence of an M-matrix of order n of type II, implies the existence of an SPBIB design with parameters

v = n = b, r = k = n/2, $\lambda_i$ values vary from 0 to $\frac{n-2}{2}$. The other parameters $n_i$ and $p_{jk}^i$ can be evaluated, in numerical problems, as it is complex to find them theoretically.

**Proof.** In the M-matrix of Type II, consider all -1's as zeros. There are n rows and n columns and hence v = b = n. In each row and each column the number of +1's is n/2. Hence r = k = n/2. And the $\lambda_i$ values vary from 0 to $\frac{n-2}{2}$ depending on the coincidences of 1's in the rows, since maximum coincidence may be $\frac{n-2}{2}$.

**Proposition 2.9.** The existence of an M-matrix of order n, of type II, implies the existence of a regular bipartite graph.

Procedure for the construction: For the matrix that we have taken, considering that as the adjacency matrix with elements as one set and the columns as another set, and the graph is defined with the condition if an element $a_{ij}$ is in column $c_k$ then ($a_{ij}$,$c_k$ ), where 1≤i,j,k≤ n, will be an edge, otherwise not. Here V = n, $V_1$= n/2 and $V_2$ = n/2 and the valence is n/2 as each row and each column has n/2 number of +1's. Hence we will get a regular bipartite graph.

**Example 2.4.** Take n = 4, then n+1 = 5 a prime. Consider the equation M = ($a_{ij}$) = (i.j) mod 5 for i, j = 1,2,3,4. We get the $M_n$-matrix as follows:

$$\begin{bmatrix} 1 & 2 & 3 & 4 \\ 2 & 4 & 1 & 3 \\ 3 & 1 & 4 & 2 \\ 4 & 3 & 2 & 1 \end{bmatrix}.$$

Now substituting 1 for even numbers and -1 for odd numbers and for 1 also in the above matrix we get the M-matrix as below

$$M = \begin{bmatrix} -1 & 1 & -1 & 1 \\ 1 & 1 & -1 & -1 \\ -1 & -1 & 1 & 1 \\ 1 & -1 & 1 & -1 \end{bmatrix}.$$

In this example n = 4, and the formula is 4k-n. We get the orthogonal number 4, the case of identical rows. And that is the case when k = 2 in the formula 4k-n. The orthogonal numbers are

$\langle R_1, R_2 \rangle = \langle R_1, R_3 \rangle = \langle R_2, R_4 \rangle = \langle R_3, R_4 \rangle = 0, \langle R_2, R_3 \rangle = \langle R_1, R_4 \rangle = -4,$ and $\langle R_i, R_i \rangle = 4, i = 1,...,n.$

This is the trivial orthogonal number. Now consider -1's are as 0's. We get the incidence matrix of a PBIB design as follows:

$$N = \begin{bmatrix} 0 & 1 & 0 & 1 \\ 1 & 1 & 0 & 0 \\ 0 & 0 & 1 & 1 \\ 1 & 0 & 1 & 0 \end{bmatrix}.$$

which is a 2-associate class GD PBIB design with parameters v = b = 4, r = k = 2, $\lambda_1 = 0$, $\lambda_2 = 1$, $n_1 = 2$, $n_2 = 1$,

$$P_1 = \begin{bmatrix} p_{11}^1 = 0 & p_{12}^1 = 1 \\ p_{21}^1 = 1 & p_{22}^1 = 0 \end{bmatrix}, P_2 = \begin{bmatrix} p_{11}^2 = 2 & p_{12}^2 = 0 \\ p_{21}^2 = 0 & p_{22}^2 = 0 \end{bmatrix}.$$

**Example 2.5.**

If we take n = 6 as n+1 = 7 a prime (as this type of matrix is being considered in example 2.2), by analogous construction we get a symmetric PBIB design with parameters v = b = 6, r = k = 3, $\lambda_1 = 2$, $\lambda_2 = 1$, $\lambda_3 = 0$, $n_1 = 2$, $n_2 = 2$, $n_3 = 1$, $p_{jk}^i$'s are given in the construction below.

**Construction:** By considering the equation (i.j) mod 7, for i, j = 1, 2, 3,4,5,6, we get the $M_n$-matrix as follows:

$$\begin{bmatrix} 123456 \\ 246135 \\ 362514 \\ 415263 \\ 531642 \\ 654321 \end{bmatrix},$$

by changing the even numbers as -1's and odd numbers as +1's and keeping 1's in the matrix as it is we get the following M-matrix.

$$M = \begin{bmatrix} 1 & -1 & 1 & -1 & 1 & -1 \\ -1 & -1 & -1 & 1 & 1 & 1 \\ 1 & -1 & -1 & 1 & 1 & -1 \\ -1 & 1 & 1 & -1 & -1 & 1 \\ 1 & 1 & 1 & -1 & -1 & -1 \\ -1 & 1 & -1 & 1 & -1 & 1 \end{bmatrix},$$

In this case the orthogonal numbers are given by

$\langle R_1, R_2 \rangle = \langle R_1, R_4 \rangle = \langle R_2, R_4 \rangle = \langle R_3, R_5 \rangle = \langle R_4, R_6 \rangle = \langle R_5, R_6 \rangle = -2$

$\langle R_1, R_3 \rangle = \langle R_1, R_5 \rangle = \langle R_2, R_3 \rangle = \langle R_2, R_6 \rangle = \langle R_4, R_5 \rangle = \langle R_4, R_6 \rangle = 2$

$\langle R_1, R_6 \rangle = \langle R_2, R_5 \rangle = \langle R_3, R_4 \rangle = -6$.  $\langle R_i, R_i \rangle = 6$, which is the trivial orthogonal number.

This type of matrix is not having the other trivial orthogonal number. The orthogonal pairs form when $\theta_1 + \theta_2 = \frac{n}{2}$. Now from the formula 4k-n it can be seen that the sum of the numbers in an orthogonal pair is 0. Here the pairs are (-2, 2), (-6, 6). As there is no row with all unities, the question of $\langle R_1, R_j \rangle$ does not arise.

by changing -1's as 0's in the above M-matrix n we get

$$N = \begin{bmatrix} 1 & 0 & 1 & 0 & 1 & 0 \\ 0 & 0 & 0 & 1 & 1 & 1 \\ 1 & 0 & 0 & 1 & 1 & 0 \\ 0 & 1 & 1 & 0 & 0 & 1 \\ 1 & 1 & 1 & 0 & 0 & 0 \\ 0 & 1 & 0 & 1 & 0 & 1 \end{bmatrix}.$$

This is the incidence matrix of an SPBIB design, with the parameters as v = b = 6, r = k = 3, $\lambda_1$= 2, $\lambda_2$ = 1, $\lambda_3$ = 0, $n_1$ = 2, $n_2$ = 2, $n_3$ = 1, and

$$P_1 = \begin{bmatrix} p^1_{11}=0 & p^1_{12}=1 & p^1_{13}=0 \\ p^1_{21}=1 & p^1_{22}=0 & p^1_{23}=1 \\ p^1_{31}=0 & p^1_{32}=1 & p^1_{33}=0 \end{bmatrix}, P_2 = \begin{bmatrix} p^2_{11}=1 & p^2_{12}=0 & p^2_{13}=1 \\ p^2_{21}=0 & p^2_{22}=1 & p^2_{23}=0 \\ p^2_{31}=1 & p^2_{32}=0 & p^2_{33}=0 \end{bmatrix}, P_3 = \begin{bmatrix} p^3_{11}=0 & p^3_{12}=2 & p^3_{13}=0 \\ p^3_{21}=2 & p^3_{22}=0 & p^3_{23}=0 \\ p^3_{31}=0 & p^3_{32}=0 & p^3_{33}=0 \end{bmatrix}.$$

Similarly by treating the above incidence matrix of the given M-matrix as an adjacent matrix of a graph we get a regular bipartite graph as follows:

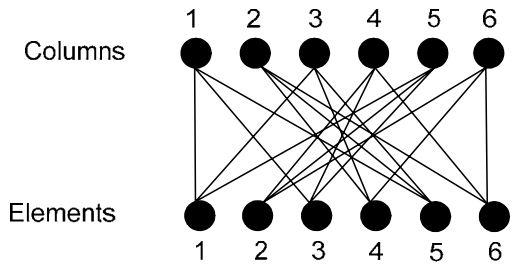

This leads to a net work problem, which can be stated as follows.

Suppose there are n points ($a_1, a_2, \ldots, a_n$) and let each point is connected to m points, m<n. By considering any one point as a source point and the remaining n-1 points as destination points find the maximum possible number of distinct routes from the source point to a destination point, we can as well impose many other restrictions on this system.

**Conclusion:** In this paper we have considered M-matrices of Type I and Type II, which are structurally different, but serve the same purpose. In Type I, we delete the first row and the first column of unities, but in Type II we retain the element 1 in the matrix. These matrices are (1,-1)-non-orthogonal matrices useful for many applications as mentioned earlier. But some problems remained open. For example some orthogonal numbers theoretically exist but do not found to be in the numerical problems. In the case of n = 11 in type I, the orthogonal numbers -5 and 7 do not exist. And the matrix $MM'$ could not be formulated and their row sums could not be established by suitable formulae in both the types. It is one way, the diagonal elements are n only and all other entries are the orthogonal numbers.

More technical features of these applications can be seen in a sequel to this paper to appear [21] shortly. And for the other details of these networks refer to Bhuyan [2], Hawkes [13], Nguyen, Vo, Lee [22].

Further work can be seen in a sequel to this paper to appear shortly.

**Acknowledgements:** One of the authors Mohan is thankful to Prof. M.G. K. Menon, who is a fountainhead of inspiration to him and to the Third World Academy of Sciences, Trieste, Italy and Prof. Bill Chen, Center for Combinatorics, Nankai University, Tianjin, PR China, for giving him an opportunity to work in the center in China for three months. His thanks are also due to Sir CRR College authorities namely K.Srimanarayana, Principal and Gutta Subbarao, Secretary for their kind support in his research quest. He is also thankful to Prof. Moon Ho Lee for extending invitation to visit Chonbuk National University, South Korea.

This work was partially supported by the MIC (Ministry of Information and Communication), under the ITFSIP (IT Foreign Specialist Inviting Program ) supervised by IITA, under ITRC supervised by IITA, and International Cooperation Research Program of the Ministry of Science & Technology, Chonbuk National University, Korea and partially by the Third World Academy of Sciences, Italy, Hence all the concerned authorities are gratefully acknowledged.